\documentclass[10pt,leqno]{article}
\usepackage{graphicx}
\usepackage{indentfirst,csquotes}

\usepackage{amssymb,amsthm,amsmath}
\usepackage{xcolor,paralist,hyperref,titlesec,fancyhdr,etoolbox}

\titleformat{\section}[display]{\normalfont\huge\bfseries\centering}{\centering\chaptertitlename\thechapter}{10pt}{\Large}
\titlespacing*{\section}{0pt}{0ex}{0ex}

\hypersetup{ colorlinks=true, linkcolor=black, filecolor=black, urlcolor=black }

\begin{document}
\title{Brightness and purity of a room-temperature single-photon source in the blue-green range} 

\author{Francis Granger{\textsuperscript{1,2,*}},Saransh Raj Gosain{\textsuperscript{1}},
Gilles Nogues{\textsuperscript{2}} \\ 
Edith Bellet-Amalric{\textsuperscript{1}}, Joël Cibert{\textsuperscript{2}}, David Ferrand{\textsuperscript{2}}, Kuntheak Kheng{\textsuperscript{1}}
}

\date{\today}
\maketitle

\let\thefootnote\relax

\begin{small}
\noindent
\textsuperscript{1} Univ. Grenoble-Alpes, CEA, Grenoble INP, IRIG, PHELIQS, NPSC, 38000 Grenoble, France \\ 
\textsuperscript{2} Univ. Grenoble-Alpes, CNRS, Inst. NEEL, 38042 Grenoble, France \\
\textsuperscript{*} Correspondence: francis.granger@cea.fr, kuntheak.kheng@cea.fr 
\end{small}

\begin{abstract}
We discuss a promising solid-state system that emits single photons at room temperature in the blue-green range, making it an attractive candidate for quantum communications in free space and underwater. The active element is a core-shell ZnSe tapered nanowire embedding a single CdSe quantum dot grown by molecular beam epitaxy. A patterned substrate enables a comprehensive study of a single nanowire using various methods. Our source shows potential for achieving a total brightness of 0.17 photon per pulse and anti-bunching with $g^{(2)}(0) < 0.3$ within a restricted spectral window. Additionally, we analyze the impact of charged excitons on the $g^{(2)}(0)$ value in different spectral ranges.
\end{abstract} 

\bigskip

Quantum communication is a rapidly growing field with enormous potential for secure data transmission. One major challenge in this field is the development of single-photon sources (SPS) that can effectively generate on-demand single photons for use in quantum information systems. The main advantage of using solid-state systems such as  quantum dots (QDs) is that they present a high purity of the emitted single-photons, making them ideal for use as flying qubits in quantum key distribution protocols \cite{arakawa_progress_2020}. In contrast to classical bits, flying qubits cannot be cloned, making it impossible for eavesdroppers to intercept the transmission secretly.

\vspace{0.3cm}

Many single-photon sources are based on III-V grown QDs and exhibit spectrally narrow lines in near infra-red when operating at cryogenic temperatures, with impressive performances in terms of brightness and purity \cite{senellart_high-performance_2017, claudon_highly_2010}. However, the need for cryogenic temperatures is still a major drawback for practically implementing these emitters in quantum information systems, despite efforts toward elevated temperatures \cite{laferriere_position-controlled_2023-1,Zeng:22,Murtaza:23}. Nitride semiconductors have shown room-temperature single-photon emission at different wavelengths  \cite{zhou_room_2018,deshpande2014,Chen2022}. CdSe QDs offer an alternative solution, as they are operational at 300K \cite{bounouar_ultrafast_2012,morozov_purifying_2023}. They emit in the blue-green range, thus allowing for free-space long-distance communication in seawater and air \cite{LI2019220}. This makes them an attractive alternative for real-world implementation of SPS in quantum information systems. Efficient light extraction from QDs is a critical issue towards high emission rates. A good way to achieve this is to embed a QD in a nanowire (NW) \cite{friedler_solid-state_2009,claudon_highly_2010,reimer_bright_2012} which acts as a single-mode waveguide channeling the photons emitted by the QD. Furthermore, a tapered NW improves the control of the emission properties. The tapered shape adiabatically expands the guided mode and reduces the divergence angle, increasing the collection efficiency \cite{gregersen_controlling_2008,jaffal_inas_2019,claudon_highly_2010,reimer_bright_2012}. 

\vspace{0.3cm}

We report here on a SPS based on a semiconductor QD-tapered NW, with purity $g^{(2)}(0)<0.3$, emitting in the blue-green range and operating at room temperature. Our study focuses on a unique QD-NW submitted to a whole range of characterizations, although other emitters on the same sample exhibit similar properties. 

\vspace{0.3cm}

\begin{figure}[t!]
\centering\includegraphics[width=0.55\linewidth]{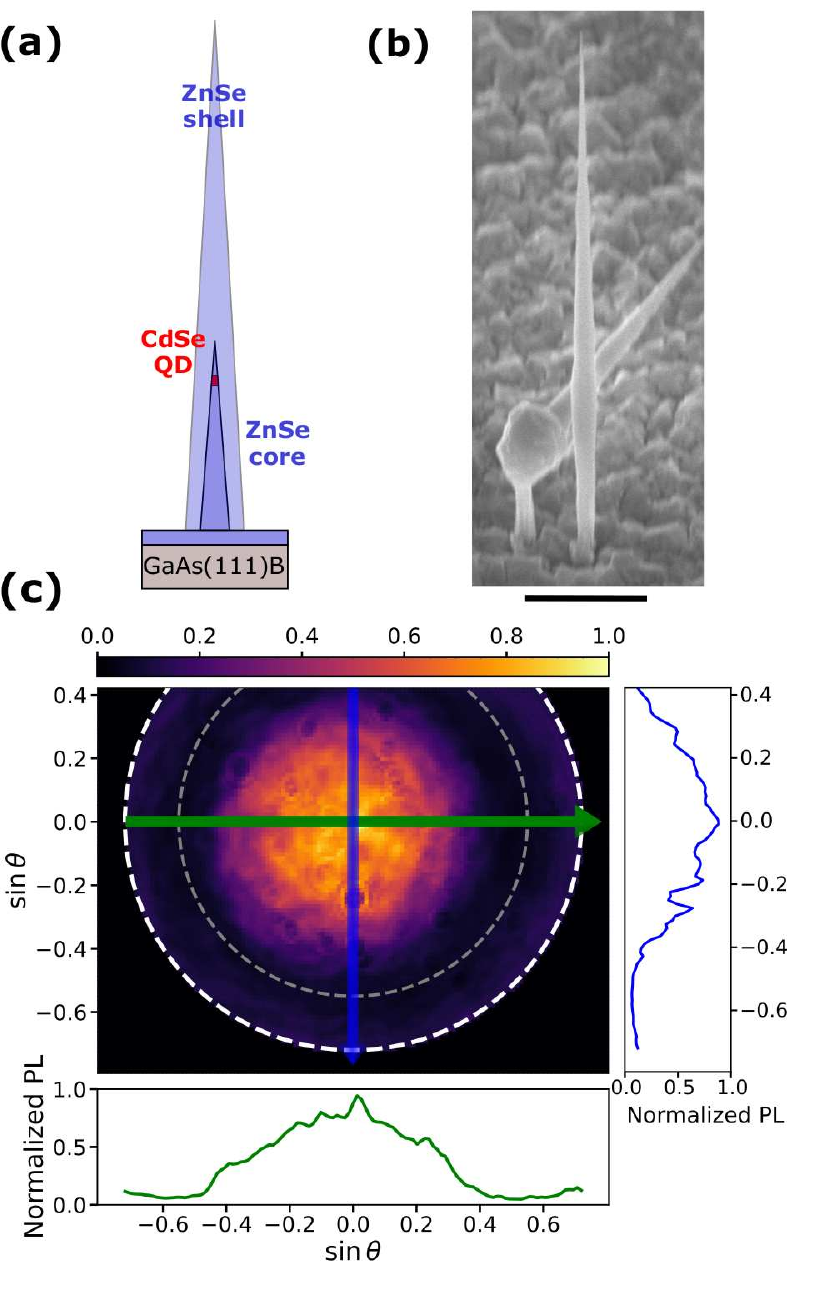}
\caption[width=\linewidth]{\textbf{(a)} Schematic of the core-shell tapered NW. \textbf{(b)} Tilted (45\textdegree C) SEM image of the studied NW (scale bar 1 \textmu m) alongside another NW that is laid down. \textbf{(c)} Radiation pattern of the same QD-NW, under 14 \textmu W  cw excitation at 405 nm. The white dashed line represents the numerical aperture $NA=0.72$ of the microscope objective used for Fourier imaging, and the gray circle is the numerical aperture $NA=0.55$ of the microscope used for spectra and TRPL measurements. The cross-sections (blue and green curves) are averaged over 5 pixels.}
\label{fig:figure1}
\end{figure}

We have used molecular beam epitaxy (MBE) to grow, at 350\textdegree C on a GaAs (111)B substrate, the ZnSe core embedding a CdSe QD near its top (Fig. \ref{fig:figure1}(a)). This was followed by the growth at 320\textdegree C of a tapered ZnSe shell. Growth conditions are detailed in Refs. \cite{gosain_onset_2022} and \cite{gosain_room_2021}. The epiready substrate was patterned before growth by combining laser lithography with chemical etching, enabling the NW position to be marked and allowing several observations on the same NW by switching from one experimental setup to another.

\vspace{0.3cm}

The vertical NW depicted in this article was selected from its far-field emission diagram recorded using Fourier imaging \cite{bulgarini_nanowire_2014}. To this purpose, we use a high numerical aperture objective ($NA=0.72$) along with a cw excitation laser at 405 nm. Then, we performed micro-photoluminescence measurements (\textmu-PL) as well as time-resolved photoluminescence (TRPL) in a second setup with non-resonant pulsed excitation from a titanium-sapphire laser  frequency-doubled  to 440 nm (repetition time $T_0 = 13.1$ ns, frequency $f_{rep} = 76$ MHz). Light is collected through a $NA=0.55$ objective and sent to a spectrometer equipped with a 600 grooves/mm grating (input slit width 0.2 mm). In order to identify the involved transitions, complementary measurements were performed at cryogenic temperature using a 1800 grooves/mm grating and a slit width of 0.05 mm. A Hanbury Brown and Twiss (HBT) setup composed of two fast avalanche photodiodes (APD) allowed us to characterize the purity of the SPS through correlation measurements. Finally, the NW was observed under a Scanning Electron Microscope (SEM), as displayed in Fig. \ref{fig:figure1}(b).

\vspace{0.3cm}

The 1 \textmu m scale bar is of the same order as the diameter of the excitation laser spot. The vertical NW has a base diameter of 140~nm and a height of 5 \textmu m. The far-field diagram in Fig. \ref{fig:figure1}(c) shows that the photons emitted upward exhibit a Gaussian mode profile with divergence angle $\alpha$ such that $\sin \alpha = 0.3$ at half maximum,  in agreement with previous studies on NWs with a small taper angle \cite{gregersen_controlling_2008,jaffal_inas_2019}. Note that the upward emission of a CdSe QD without the thick tapered ZnSe shell is very weak due to the dielectric screening present in nanowires with small diameters \cite{jeannin_enhanced_2017}. The tapered shell ensures an adiabatic coupling between the guided mode and free space, leading to a collection efficiency within $NA=0.72$ calculated to be close to unity. Measurements in Fig.~\ref{fig:figure1}(c) show that 85\% of the collected light falls within $NA = 0.55$. We note that light emitted towards the substrate is lost, leading to a theoretical maximum source efficiency close to 50\%. Additionally, the microscope objective with $NA = 0.55$ shows a transmission of 90\% at the wavelength of interest \cite{Mitutoyo}.  Thus, an ideal SPS made of a QD inserted in the present NW would provide a photon flux around 0.38 photon per pulse at the output of this objective.

\vspace{0.3cm}

\begin{figure}[ht!]
\centering\includegraphics[width=0.6\linewidth]{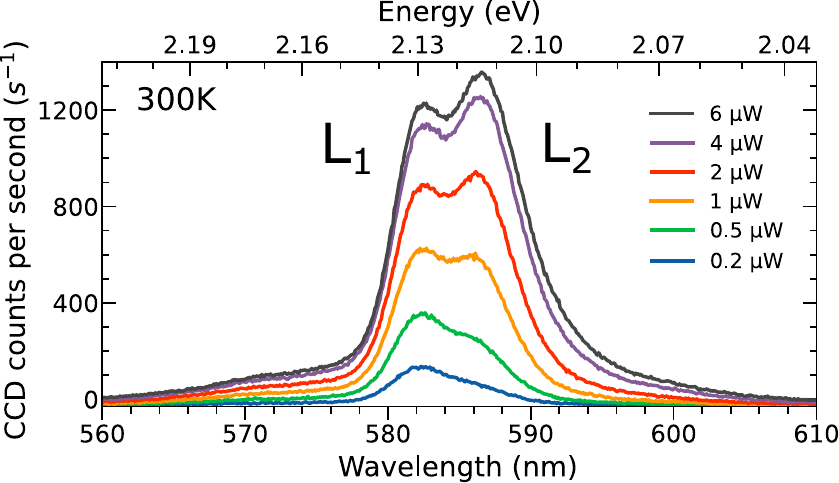}
\caption{
Power-dependent spectra of the QD-NW of Fig. \ref{fig:figure1}(b).
}
\label{fig:figure2}
\end{figure}

The QD-NW room temperature \textmu-PL spectra are shown in Fig.~\ref{fig:figure2} for values of the pulsed excitation power $P$ ranging from 0.2 to 6~\textmu W. We observe two  peaks at 582.5~nm (L1) and 586.5~nm (L2). It can be seen qualitatively that the L1 intensity grows linearly with $P$, while L2 grows at a faster rate. At $P=6$~\textmu W, the QD starts to saturate. Using the setup calibration  explained in Supplement 1, the number of photons collected out of the microscope objective for L1 and L2 together yield to a promising value of 0.28 photon per pulse. This is discussed later on.

\vspace{0.3cm}

\begin{figure}[ht!]
\centering
\includegraphics[width=0.55\linewidth]{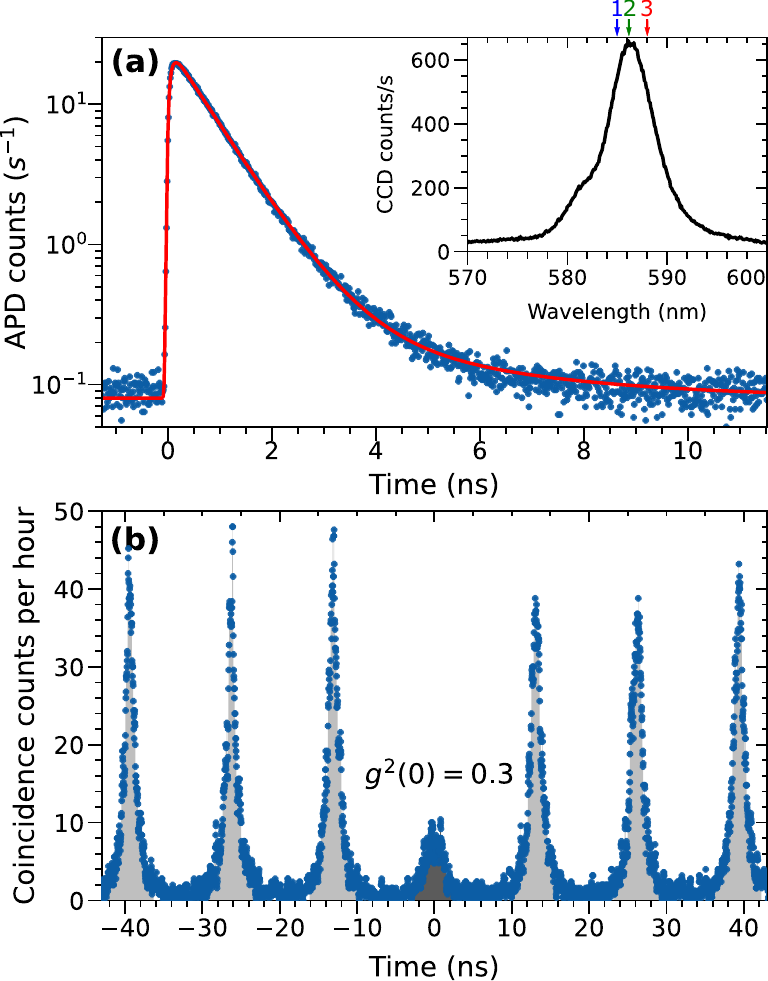}
\caption{Room temperature time-resolved measurements on the same NW as in Fig. \ref{fig:figure1}(b). \textbf{(a)} TRPL decay at $585 \pm 0.5$ nm ($P=5.3$~~\textmu W). The solid red line is a fit using Eq.~\ref{eq:fit}. The inset displays the µ-PL spectrum at 5.3~\textmu W, with arrows indicating the central positions of spectral bandwidths (1 nm) used for time-resolved measurements: $585 \pm 0.5$ nm (blue arrow 1), $586.2 \pm 0.5$ nm (green arrow 2), and $588 \pm 0.5$ nm (red arrow 3). \textbf{(b)} Second-order intensity correlation of the QD emission under 5~\textmu W pulsed excitation. The symbols show the histogram of delay times between  photons detected on the start-stop APDs, averaged over seven channels. The value $g^{(2)}(0) = 0.3$ has been calculated by removing a constant baseline and computing the ratio of the central peak area (dark grey) to the lateral ones (light grey).}
\label{fig:figure3}
\end{figure}

The onset of saturation around $P=6$~\textmu W was confirmed by further measurements up to 20~\textmu W. However, these measurements at high-power  caused an evolution of the spectrum, as seen in the inset of Fig. \ref{fig:figure3}(a) for $P=5.3$~\textmu W. The L1 intensity decreased significantly and L2 to a lesser extent. After this evolution, L1 + L2 signals yield to 0.09 photon per pulse, a value which remained constant for weeks after this initial decrease. 

\vspace{0.3cm}

Figure \ref{fig:figure3} shows the TRPL signal of the QD-NW emission within a spectral window of $585 \pm 0.5$~nm (blue arrow labeled 1 in the inset). The red solid line in Fig. \ref{fig:figure3}(a) is the convolution of the response function of the setup (a Gaussian with a FWHM of 60~ps~\cite{noauthor_tcspc_nodate}), and a sum of three exponential functions added to a constant baseline:

\begin{equation}\label{eq:fit}
    I(t) = \sum_{i=1}^{3} A_i\exp(-\frac{t}{\tau_i}) + B
\end{equation}

 The main component is a decay with a short lifetime $\tau_2 = 0.74$~ns. There is also a weak, slow component, $A_3\exp(-\frac{t}{\tau_3})$ with $\tau_3 = 3$ ns and $A_3/A_2 < 0.06$. The term $A_1\exp(-\frac{t}{\tau_1})$, introduced to describe a population rise of the emitting state, was found to be hidden by the time-resolution of the APD \cite{noauthor_tcspc_nodate}. 

The purity of the SPS was determined using a HBT autocorrelation measurement. The coincidence histogram for the same conditions as in Fig.~\ref{fig:figure3}(a) and the same spectral window (1 nm centered on the blue arrow) is shown in Fig. \ref{fig:figure3}(b). Under pulsed excitation, a series of correlation peaks appear, with equal spacing given by the laser repetition period $T_0$. The peak widths agree with the lifetime $\tau_2$. As $\tau_2 \ll T_0$, the purity is computed using the ratio of the central peak area to the lateral ones. At 585.0 nm, our system exhibits a second-order correlation at zero delays $g^{(2)}(0) = 0.3$, below the threshold of 0.5 considered as a signature of single-photon emission. 

\vspace{0.3cm}

\begin{figure}[ht!]
\centering\includegraphics{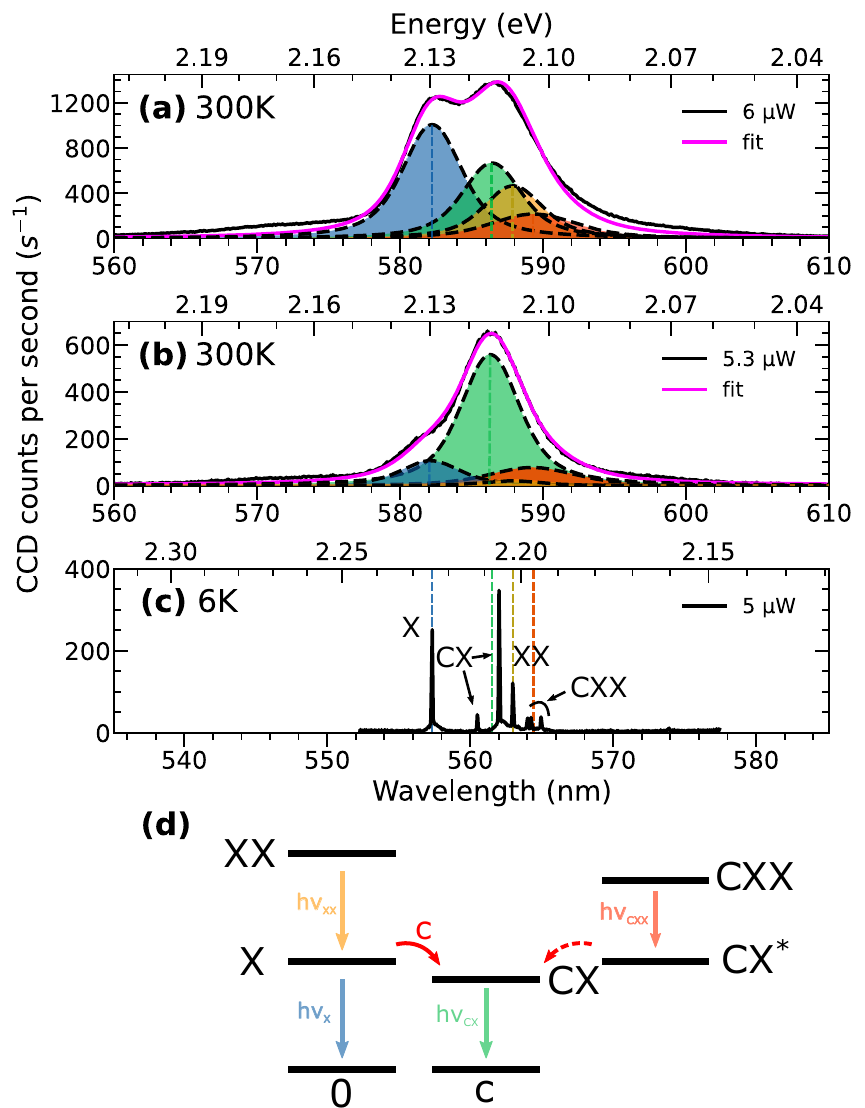}
\caption{
PL measurements on the same NW as in Fig. \ref{fig:figure1}(b). \textbf{(a)} Spectrum under 6 \textmu W excitation at room temperature. The signal is fitted (magenta curve) with a sum of four Voigt functions, resulting in the blue, green, yellow and red filled-lines. The same method is applied for the spectrum after intensity decrease in \textbf{(b)}. \textbf{(c)} Spectrum performed at 6K, under 5 \textmu W pulsed excitation, with grating 1800 grooves/mm, slits 0.05 mm. The wavelength range was shifted to match the X lines with the 300K spectrum. In all the spectra, constant baselines have been removed (electronic noise).
\textbf{(d)} Schematic diagram of the mechanisms possibly involved in the luminescence process: the simple radiative cascade (neutral or charged) and a possible cross-radiative cascade, involving CX resulting for the transfer of an extra charge.}
\label{fig:figure4}
\end{figure}

It is difficult to identify the nature of L1 and L2 from the spectra at room temperature, but this can be done by a comparison with the PL spectrum at 6K shown in Fig.~\ref{fig:figure4}(c). Sharp zero-phonon lines lie on top of broad acoustic-phonon sidebands. Polarization-resolved PL spectra (see Supplement 1) allow us to observe a fine structure splitting with opposite signs for two of those lines, therefore attributed to the neutral exciton X and biexciton XX. This agrees with the linear and quadratic behavior observed in Fig. \ref{fig:figure2}. The most intense line in Fig.~\ref{fig:figure4}(c) shows no fine structure. Hence, it is attributed to a charged exciton CX. A similar spectral arrangement of the X, CX, and XX lines has been observed previously on similar samples \cite{gosain_quantitative_2022}. The CX line was attributed to the positively charged exciton $X^+$ in Ref.~\cite{bounouar_photon_2012}, and the lines with energies smaller than XX were attributed to the charged biexciton CXX in Refs.~\cite{bounouar_photon_2012,akimov_fine_2002}. Finally, the small line at 560.8~nm in Fig. \ref{fig:figure4}(c) was attributed to the negatively charged exciton X$^{-}$ in \cite{jeannin_enhanced_2017}. We note a large splitting of 5.66~nm (22.4 meV) between the X and XX lines.

\vspace{0.3cm}

At room temperature, the zero-phonon lines are vanishingly small, and the spectra only show broad phonon sidebands. The spectra for $P=6$~\textmu W in Fig. \ref{fig:figure2} and the inset of Fig.\ref{fig:figure3}(a) are fitted and displayed in Fig.\ref{fig:figure4}(a) and (b), respectively. They comprise four Voigt functions and a constant baseline. The position of each line is set to match the relative positions of the X, CX, XX, and CXX lines in Fig.~\ref{fig:figure4}(c). The X, CX, and XX linewidths are constrained to a common value, while CXX, composed of several lines, has a larger width. Assuming that the trapping of electron-hole pairs does not depend on the presence of charges in the QD, the amplitude of each line is set so that the ratios of intensities obey $I_{XX}/I_X = I_{CXX}/I_{CX}$. 

\vspace{0.3cm}

The fit in Fig. \ref{fig:figure4}(b) results in a X, CX, and XX linewidth equal to 5.2 nm (19 meV), and 7.2 nm (26 meV) for CXX. The common biexciton/exciton intensity ratio is 0.45, illustrating that we are not fully at saturation at $P=6$~\textmu W. Although X is dominant in Fig. \ref{fig:figure4} (a), the contribution of CX is sizable with $I_{CX}/(I_{CX}+I_X) = 0.40$. The lines X and CX together yield 0.17 photon per pulse.

\vspace{0.3cm}

The ratios $I_{XX}/I_X = I_{CXX}/I_{CX} = 0.28$ in Fig. \ref{fig:figure4} (b) indicate that either the spectrum (after the intensity decrease) is further away from saturation, or that non-radiative processes reduce the X emission intensity. Additionally, the ratio $I_{CX}/(I_{CX}+I_X) = 0.82$ shows that CX dominates X. In Fig. \ref{fig:figure4} (b), the X and CX lines together contribute for 0.06 photon per pulse. Figure \ref{fig:figure3}(a) displays the decay at 585.0 nm, where the fit shows that the strongest contribution is due to the CX line, with a small overlap with CXX and X lines. The corresponding decay curve (Fig \ref{fig:figure3}(a)) exhibits a dominant exponential decay with a sub-nanosecond lifetime ($\tau_2 = 0.74$ ns), a value slightly larger than the lifetime reported in Ref. \cite{gosain_room_2021}. Moreover, the fast rise is in good agreement with an excitation power far from saturation.

\vspace{0.3cm}

We also investigated the impact of the spectral window position on the emitter's purity. At 585.0~nm (blue arrow in the inset of Fig.~\ref{fig:figure3}(a)), the  zero-delay correlation peak is $g^{(2)}(0)=0.3$. We measure $g^{(2)}(0)=0.4$ close to L2 peak center (green arrow labeled as 2 in the inset) and $g^{(2)}(0)=0.6$ on the red side of L2 (red arrow labeled as 3). See Supplement 1 for additional data. Each of these 1nm-broad portions of the spectrum yields to approximately 0.02 photon per pulse. We discuss below how higher values can be achieved. Shifting from spectral window 1 to 3, the overlap between CX and CXX increases, enhancing a correlation through the biexciton-exciton cascade in the charged QD. Coming back to spectral window 1, there is still some CX-CXX overlap at 585.0 nm but we cannot rule out the possibility of a cross radiative cascade resulting from a fast charge trapping forming the positively charged exciton $X + h \xrightarrow{} X^+$, as illustrated in Fig. \ref{fig:figure4}(d). We may expect to increase the purity even more ($g^{(2)}(0)<0.3$) by blue-shifting the spectral window below 585 nm. This would make it possible to identify the role of the cross-cascade.  However, the intensity diminishes drastically as we shift the spectral window to lower wavelengths. There is an unavoidable tradeoff between purity which requires a narrow spectral window, and brightness, where a broad spectral bandwidth with high intensity is needed.

\vspace{0.3cm}

In conclusion, we demonstrate a promising triggered SPS operational at room temperature, with large X-XX splitting of 5.66 nm (22.4 meV), enabling us to obtain a high single-photon purity with $g^{(2)}(0)<0.3$. We identify neutralizing the quantum dot as a highly promising way to enhance the performance of the source. Adding the full intensity of the X and CX lines, we obtain a SPS brightness ranging from 0.17 photon per pulse initially (i.e. an emission rate of 13 MHz with a 76 MHz excitation rate) to 0.06 (i.e. 4.5 MHz) after intensity decrease. The observed subnanosecond decay time would allow a much higher repetition rate towards the GHz range. Further improvements are necessary to address the observed charging effect. Additional non-resonant laser excitation may be used to fill the traps surrounding the QD. Alternatively, modifying the growth conditions \cite{jeannin_enhanced_2017} and surface passivation of the NW may also effectively mitigate surface states from serving as traps for charge carriers. Lower temperatures down to 220K (achieved using a Peltier cooler) can be used to reduce the linewidths and hence the overlap between lines undergoing a radiative cascade \cite{laferriere_position-controlled_2023-1}. We did carry out cryogenic measurements on the studied nanowire and checked that the value of $g^{(2)}(0)$ is below a few percent (see Supplement 1). Intermediate temperature measurements are underway to understand the phenomena better, and further optimize the SPS performances. The efficient coupling between the guided mode in the NW and free-space paves the way for a deterministic integration of the SPS into a photonic device \cite{mnaymneh_-chip_2020}.

\vspace{1cm}

\section*{Acknowledgments}
\vspace{0.5cm}
The substrate has been patterned by lithography at the NanoFab cleanroom facility of Institut Néel, Grenoble. Thanks are due to Yann Genuist for his participation to the growth of samples. The authors thank Jean-Michel Gérard for fruitful discussions.

\vspace{1cm}

\section*{Funding}
\vspace{0.5cm}
We acknowledge funding from the Laboratoire d \textquotesingle excellence LANEF in Grenoble, ANR-10-LABX-51-01, and CEA-PE Bottom-Up QPhotonics.

\vspace{1cm}

\section*{Disclosures}
\vspace{0.5cm}
The authors declare no conflicts of interest.

\vspace{1cm}

\section*{Data Availability Statement} 
\vspace{0.5cm}
Data may be obtained from the authors upon reasonable request.

\vspace{1cm}

$\,$

$\,$

\renewcommand\refname{References}
\begin{footnotesize}
\bibliographystyle{unsrt.bst} 
\textnormal{\bibliography{sample.bib}}

\begin{thebibliography}{10}

\bibitem{arakawa_progress_2020}
Yasuhiko Arakawa and Mark~J. Holmes.
\newblock Progress in quantum-dot single photon sources for quantum information
  technologies: {A} broad spectrum overview.
\newblock {\em Applied Physics Reviews}, 7(2):021309, June 2020.

\bibitem{senellart_high-performance_2017}
Pascale Senellart, Glenn Solomon, and Andrew White.
\newblock High-performance semiconductor quantum-dot single-photon sources.
\newblock {\em Nature Nanotechnology}, 12(11):1026--1039, November 2017.

\bibitem{claudon_highly_2010}
Julien Claudon, Joël Bleuse, Nitin~Singh Malik, Maela Bazin, Périne
  Jaffrennou, Niels Gregersen, Christophe Sauvan, Philippe Lalanne, and
  Jean-Michel Gérard.
\newblock A highly efficient single-photon source based on a quantum dot in a
  photonic nanowire.
\newblock {\em Nature Photonics}, 4(3):174--177, March 2010.

\bibitem{laferriere_position-controlled_2023-1}
Patrick Laferriére, Sofiane Haffouz, David~B. Northeast, Philip~J. Poole,
  Robin~L. Williams, and Dan Dalacu.
\newblock Position-{Controlled} {Telecom} {Single} {Photon} {Emitters}
  {Operating} at {Elevated} {Temperatures}.
\newblock {\em Nano Letters}, 23(3):962--968, February 2023.

\bibitem{Zeng:22}
Helen Zhi~Jie Zeng, Minh Anh~Phan Ngyuen, Xiaoyu Ai, Adam Bennet, Alexander~S.
  Solntsev, Arne Laucht, Ali Al-Juboori, Milos Toth, Richard~P. Mildren, Robert
  Malaney, and Igor Aharonovich.
\newblock Integrated room temperature single-photon source for quantum key
  distribution.
\newblock {\em Opt. Lett.}, 47(7):1673--1676, Apr 2022.

\bibitem{Murtaza:23}
Ghulam Murtaza, Maja Colautti, Michael Hilke, Pietro Lombardi,
  Francesco~Saverio Cataliotti, Alessandro Zavatta, Davide Bacco, and Costanza
  Toninelli.
\newblock Efficient room-temperature molecular single-photon sources for
  quantum key distribution.
\newblock {\em Opt. Express}, 31(6):9437--9447, Mar 2023.

\bibitem{zhou_room_2018}
Yu~Zhou, Ziyu Wang, Abdullah Rasmita, Sejeong Kim, Amanuel Berhane, Zoltán
  Bodrog, Giorgio Adamo, Adam Gali, Igor Aharonovich, and Wei-bo Gao.
\newblock Room temperature solid-state quantum emitters in the telecom range.
\newblock {\em Science Advances}, 4(3):eaar3580, March 2018.

\bibitem{deshpande2014}
Saniya Deshpande, Thomas Frost, Arnab Hazari, and Pallab Bhattacharya.
\newblock Electrically pumped single-photon emission at room temperature from a
  single ingan/gan quantum dot.
\newblock {\em Applied Physics Letters}, 105(14):141109, 2014.

\bibitem{Chen2022}
Ling Chen, Bowen Sheng, Shanshan Sheng, Ping Wang, Xiaoxiao Sun, Duo Li, Tao
  Wang, Renchun Tao, Shangfeng Liu, Zhaoying Chen, Weikun Ge, Bo~Shen, and
  Xinqiang Wang.
\newblock Room temperature triggered single photon emission from self-assembled
  gan/aln quantum dot in nanowire.
\newblock {\em Advanced Functional Materials}, 32(47):2208340, 2022.

\bibitem{bounouar_ultrafast_2012}
S.~Bounouar, M.~Elouneg-Jamroz, M.~den Hertog, C.~Morchutt, E.~Bellet-Amalric,
  R.~André, C.~Bougerol, Y.~Genuist, J.-Ph. Poizat, S.~Tatarenko, and
  K.~Kheng.
\newblock Ultrafast {Room} {Temperature} {Single}-{Photon} {Source} from
  {Nanowire}-{Quantum} {Dots}.
\newblock {\em Nano Letters}, 12(6):2977--2981, June 2012.

\bibitem{morozov_purifying_2023}
Sergii Morozov, Stefano Vezzoli, Alina Myslovska, Alessio~Di Giacomo,
  N.~Asger Mortensen, Iwan Moreels, and Riccardo Sapienza.
\newblock Purifying single photon emission from giant shell {CdSe}/{CdS}
  quantum dots at room temperature.
\newblock {\em Nanoscale}, 15(4):1645--1651, 2023.

\bibitem{LI2019220}
Dong-Dong Li, Qi~Shen, Wei Chen, Yang Li, Xuan Han, Kui-Xing Yang, Yu~Xu, Jin
  Lin, Chao-Ze Wang, Hai-Lin Yong, Wei-Yue Liu, Yuan Cao, Juan Yin, Sheng-Kai
  Liao, and Ji-Gang Ren.
\newblock Proof-of-principle demonstration of quantum key distribution with
  seawater channel: towards space-to-underwater quantum communication.
\newblock {\em Optics Communications}, 452:220--226, 2019.

\bibitem{friedler_solid-state_2009}
I.~Friedler, C.~Sauvan, J.~P. Hugonin, P.~Lalanne, J.~Claudon, and J.~M.
  Gérard.
\newblock Solid-state single photon sources: the nanowire antenna.
\newblock {\em Optics Express}, 17(4):2095--2110, February 2009.

\bibitem{reimer_bright_2012}
Michael~E. Reimer, Gabriele Bulgarini, Nika Akopian, Moïra Hocevar,
  Maaike~Bouwes Bavinck, Marcel~A. Verheijen, Erik P. A.~M. Bakkers, Leo~P.
  Kouwenhoven, and Val Zwiller.
\newblock Bright single-photon sources in bottom-up tailored nanowires.
\newblock {\em Nature Communications}, 3(1):737, March 2012.

\bibitem{gregersen_controlling_2008}
Niels Gregersen, Torben~R. Nielsen, Julien Claudon, Jean-Michel Gérard, and
  Jesper Mørk.
\newblock Controlling the emission profile of a nanowire with a conical taper.
\newblock {\em Optics Letters}, 33(15):1693--1695, August 2008.

\bibitem{jaffal_inas_2019}
Ali Jaffal, Walid Redjem, Philippe Regreny, Hai Son Nguyen, Sébastien Cueff,
  Xavier Letartre, Gilles Patriarche, Emmanuel Rousseau, Guillaume Cassabois,
  Michel Gendry, and Nicolas Chauvin.
\newblock {InAs} quantum dot in a needlelike tapered {InP} nanowire: a telecom
  band single photon source monolithically grown on silicon.
\newblock {\em Nanoscale}, 11(45):21847--21855, 2019.

\bibitem{gosain_onset_2022}
Saransh~Raj Gosain, Edith Bellet-Amalric, Martien~den Hertog, Régis André,
  and Joël Cibert.
\newblock The onset of tapering in the early stage of growth of a nanowire.
\newblock {\em Nanotechnology}, 33(25):255601, April 2022.

\bibitem{gosain_room_2021}
Saransh~Raj Gosain.
\newblock {\em Room temperature single-photon source based on semiconductor
  quantum-dot nanowire for integrated photonics}.
\newblock PhD thesis, Université Grenoble Alpes, 2021.

\bibitem{bulgarini_nanowire_2014}
Gabriele Bulgarini, Michael~E. Reimer, Maaike Bouwes~Bavinck, Klaus~D. Jöns,
  Dan Dalacu, Philip~J. Poole, Erik P. A.~M. Bakkers, and Val Zwiller.
\newblock Nanowire {Waveguides} {Launching} {Single} {Photons} in a {Gaussian}
  {Mode} for {Ideal} {Fiber} {Coupling}.
\newblock {\em Nano Letters}, 14(7):4102--4106, July 2014.

\bibitem{jeannin_enhanced_2017}
Mathieu Jeannin, Thibault Cremel, Teppo Häyrynen, Niels Gregersen, Edith
  Bellet-Amalric, Gilles Nogues, and Kuntheak Kheng.
\newblock Enhanced {Photon} {Extraction} from a {Nanowire} {Quantum} {Dot}
  {Using} a {Bottom}-{Up} {Photonic} {Shell}.
\newblock {\em Physical Review Applied}, 8(5):054022, November 2017.

\bibitem{Mitutoyo}
Mitutoyo corporation, 20-1, Sakado 1-Chome, Takatsu-ku, Kawasaki-shi, Kanagawa
  213-8533, Japan.
\newblock {\em Microscope units and objectives}.

\bibitem{noauthor_tcspc_nodate}
{TCSPC} {Performance} of the id100-50 detector, {Becker} and {Hickl} {GmbH}.
\newblock
  \url{https://www.photonicsolutions.co.uk/upfiles/id100-50-becker.pdf}.

\bibitem{gosain_quantitative_2022}
Saransh~Raj Gosain, Edith Bellet-Amalric, Eric Robin, Martien Den~Hertog,
  Gilles Nogues, Joël Cibert, Kuntheak Kheng, and David Ferrand.
\newblock Quantitative analysis of the blue-green single-photon emission from a
  quantum dot in a thick tapered nanowire.
\newblock {\em Physical Review B}, 106(23):235301, December 2022.

\bibitem{bounouar_photon_2012}
Samir Bounouar.
\newblock {\em Photon correlations on a room temperature semi-conductor single
  photon emitter.}
\newblock phdthesis, Université de Grenoble, February 2012.

\bibitem{akimov_fine_2002}
I.~A. Akimov, A.~Hundt, T.~Flissikowski, and F.~Henneberger.
\newblock Fine structure of the trion triplet state in a single self-assembled
  semiconductor quantum dot.
\newblock {\em Applied Physics Letters}, 81(25):4730--4732, December 2002.

\bibitem{mnaymneh_-chip_2020}
Khaled Mnaymneh, Dan Dalacu, Joseph McKee, Jean Lapointe, Sofiane Haffouz,
  John~F. Weber, David~B. Northeast, Philip~J. Poole, Geof~C. Aers, and
  Robin~L. Williams.
\newblock On-{Chip} {Integration} of {Single} {Photon} {Sources} via
  {Evanescent} {Coupling} of {Tapered} {Nanowires} to {SiN} {Waveguides}.
\newblock {\em Advanced Quantum Technologies}, 3(2):1900021, 2020.
\newblock reprint:
  https://onlinelibrary.wiley.com/doi/pdf/10.1002/qute.201900021.

\end{thebibliography}
\end{footnotesize}
\newpage

\pagebreak

\begin{center}
\textbf{\Huge Brightness and purity of a room-temperature
single-photon source in the blue-green range: supplemental document}
\end{center}

\vspace*{1cm}

\section*{1 - Calibration of the experimental setup}

\begin{figure}[htbp]
\centering
\includegraphics[width=1\linewidth]{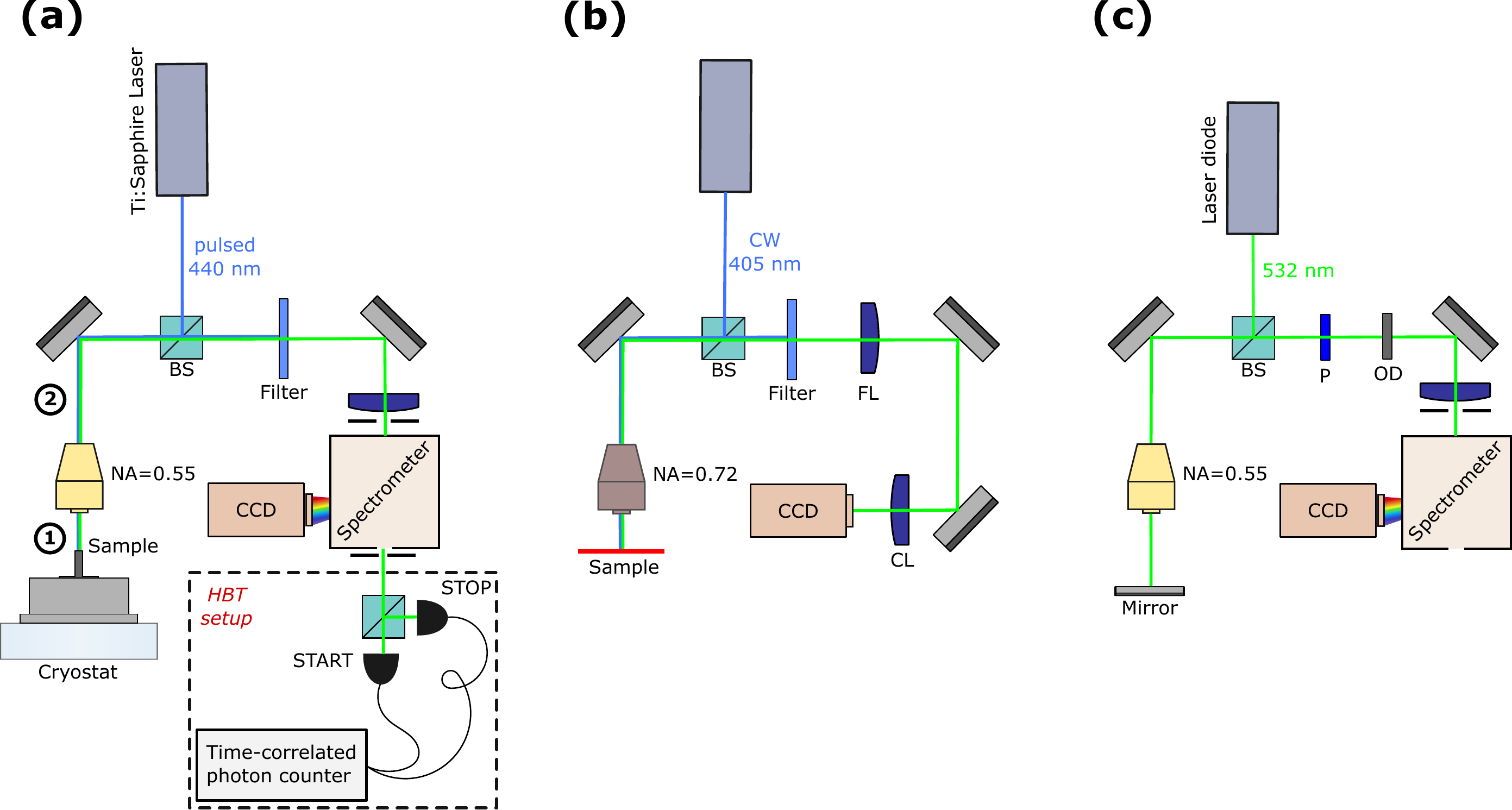}
\caption{Schematics of the spectroscopy setups. \textbf{(a)} TRPL setup. BS: beam-splitter; \textbf{(b)} Setup used for Fourier imaging technique. FL: Fourier Lens; CL: Collection Lens. \textbf{(c)} Schematic of the setup calibration. OD: Optical Density; P: Polarizer}
\label{fig:setup}
\end{figure}

\subsection*{Experimental setup}
\vspace{0.5cm}
The setup used for micro-PL and TRPL measurements presented in the article (at 6K and 300K) is displayed in Figure \ref{fig:setup} (a). The sample was positioned on a continuous flow cold-finger cryostat, which can cool the sample down to 5-6 K for cryogenic measurements. The cryostat is mounted on a X-Y linear stage, enabling high precision and smooth motion control. The excitation is provided by a frequency doubled pulsed Ti:Sapphire laser at 440 nm. This results in pulses (< 2 ps) with a repetition frequency of $f_{\text{rep}} = 76$ MHz that excite the NW along its axis. A $NA=0.55$ M Plan apo SL 100X Mitutoyo microscope objective is used to focus the excitation laser on a selected NW-QD and to collect the QD emitted light. Emission spectra are recorded using a Horiba Jobin Yvon HR-460 spectrometer and a DU920P-BEX2-DD Andor CCD camera. Second-order correlation measurements are performed on an HBT setup equipped with two ID Quantique single-photon detectors (ID100-50) mounted on the side exit of the spectrometer. The correlations are computed using a SPC-130 Becker and Hickl time-correlated photon counting module (TCSPC).

\vspace{0.3cm}

For the Fourier imaging technique, a second setup (Figure \ref{fig:setup} (b)) is used, which extends the confocal microscopy setup with an additional lens (Fourier lens in Figure \ref{fig:setup} (b)). The lens is used to project the back focal plane of the objective (image plane) onto a CCD camera, enabling Fourier imaging of the emitter. In the present 2-lens configuration, given that the Fourier lens (FL) and convergence lens (CL) have long focal lengths of 400 mm and 250 mm, we do not expect a significant impact on the Fourier plane image quality compared to a 4f lens configuration.

\subsection*{Setup calibration and emission line brightness}
\vspace{0.5cm}
In this paper, we define the brightness of an emission line ($B_{\text{line}}$) as the average photon number transmitted by the microscope objective for each laser excitation pulse (\emph{i.e} the averaged photon number per pulse measured at the position 2 of Figure \ref{fig:setup}.a). Note that the brightness is sometimes defined in the literature as the photon number collected by the microscope objective (\emph{i.e} the averaged photon number per pulse measured at the position 1 of Figure \ref{fig:setup}.a). At the QD emission wavelength, about 10\% of QD emission is lost in the microscope objective and $B_{line}$ represents an effective brightness that can be used for practical implementations. For a pulsed laser excitation with the repetition frequency $f_{\text{rep}}$, $B_{\text{line}}$ is directly related to the line emission rate $\gamma_{\text{line}}$ measured at the position 2: 

\begin{equation}
    B_{\text{line}}= \frac{\gamma_{\text{line}}}{f_{\text{rep}}}
\end{equation}

For the measurements done with the set-up shown in Figure \ref{fig:setup}.(a), we compute the line emission rate $\gamma_{\text{line}}$ from the integrated count rate ($\gamma_{\text{CCD}}$) of the CCD spectra recorded under pulsed excitation : $\gamma_{\text{CCD}}= \epsilon \gamma_{\text{line}}$ where $\epsilon$ is the set-up collection efficiency.\\

To evaluate $\epsilon$ at the QD emission wavelength (580 to 590 nm), we use an attenuated 532 nm cw laser diode (see Figure \ref{fig:setup}.c). The value of $\epsilon$ was deduced from the ratio of the laser diode integrated CCD spectra and the laser power measured in front of the optical density (OD=4). At this particular wavelength, we observe that $\epsilon$ does not depend on the incident polarization. The influence of the injection beam splitter,  the spectrometer entrance slit and of the spot size was taken into account. The transmission of the 50/50 injection beam splitter was checked at the laser diode wavelength. Please note that the CCD camera was used with an internal gain of 10 (one CCD count corresponding to 10 detected photons). In our setup, $\epsilon$ was evaluated to be approximately $9.2 \times 10^{-3}$ using a 0.2 mm entrance slit.

\section*{2 - Additional data}

\vspace{1cm}

\subsection*{Room-temperature TRPL measurements}
\vspace{0.5cm}
\begin{figure}[htbp]
\centering
\includegraphics[width=0.9\linewidth]{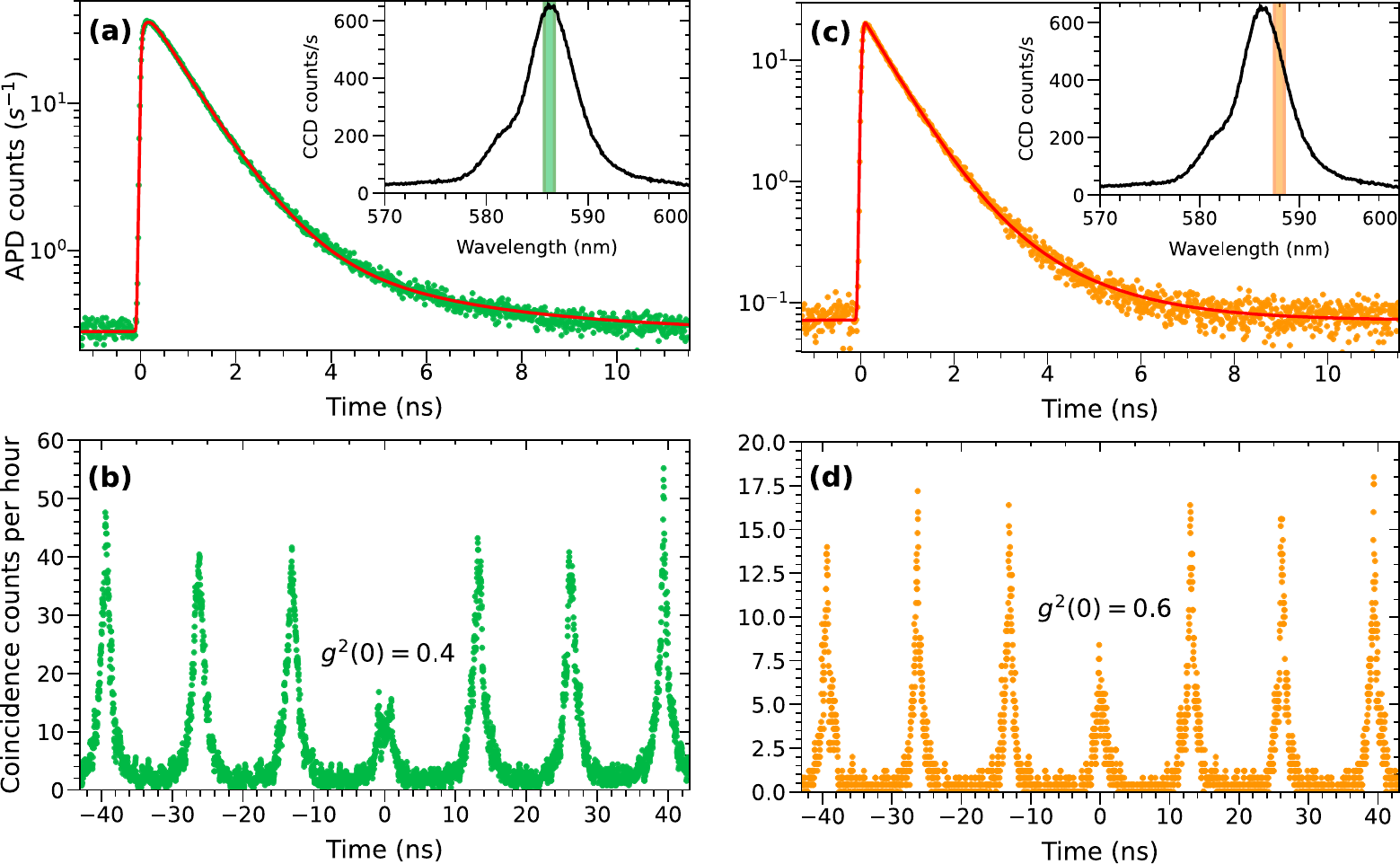}
\caption{Room temperature TRPL measurements on the NW studied through the article. \textbf{(a)} Decay time of the QD-NW under 5~\textmu W excitation power at $586.2 \pm 0.5$ nm (green symbols) and $588.0 \pm 0.5$ nm in \textbf{(c)} (orange symbols). An exponential decay model is used to fit the signals (red curves). \textbf{(b)} Second-order intensity correlation function at 4.8~\textmu W. The data are averaged over seven channels. The traces exhibit $g^2(0) = 0.4$ at 586.2 nm (green symbols), and in \textbf{(d)} $g^2(0) = 0.6$ at 588.0 nm (orange symbols).}
\label{fig:TRPL}
\end{figure}

Additional TRPL data are presented in Figure \ref{fig:TRPL}, for two spectral window positions (centered at 586.2 nm and 588 nm).
Table \ref{tab:shape-functions} shows the parameters used to fit the decay time curves shown in Figure \ref{fig:TRPL} (a) and (c) insets. 

\begin{table}[htbp]
\centering
\caption{\bf Values of parameters used in Figure \ref{fig:TRPL} (a) and (c)}
\begin{tabular}{cllc}
\hline
Window central position (nm) & Contribution & $A_i$ & $\tau_i$ (ns)\\
\hline
586.2 & Rise (i = 1)       & -2.76 & 0.12 \\
      & Fast decay (i = 2) & 39.85 & 0.81\\
      & Slow decay (i = 3) & 4.16 & 3.07\\
588.0   & Rise (i = 1)       & 0           &  \\
      & Fast decay (i = 2) & 13.93        & 0.64\\
      & Slow decay (i = 3) & 2.41        & 1.65\\
\hline
\end{tabular}
  \label{tab:shape-functions}
\end{table}

\subsection*{Fine-structure splitting}
\begin{figure}[htbp]
\centering
\includegraphics[width=0.7\linewidth]{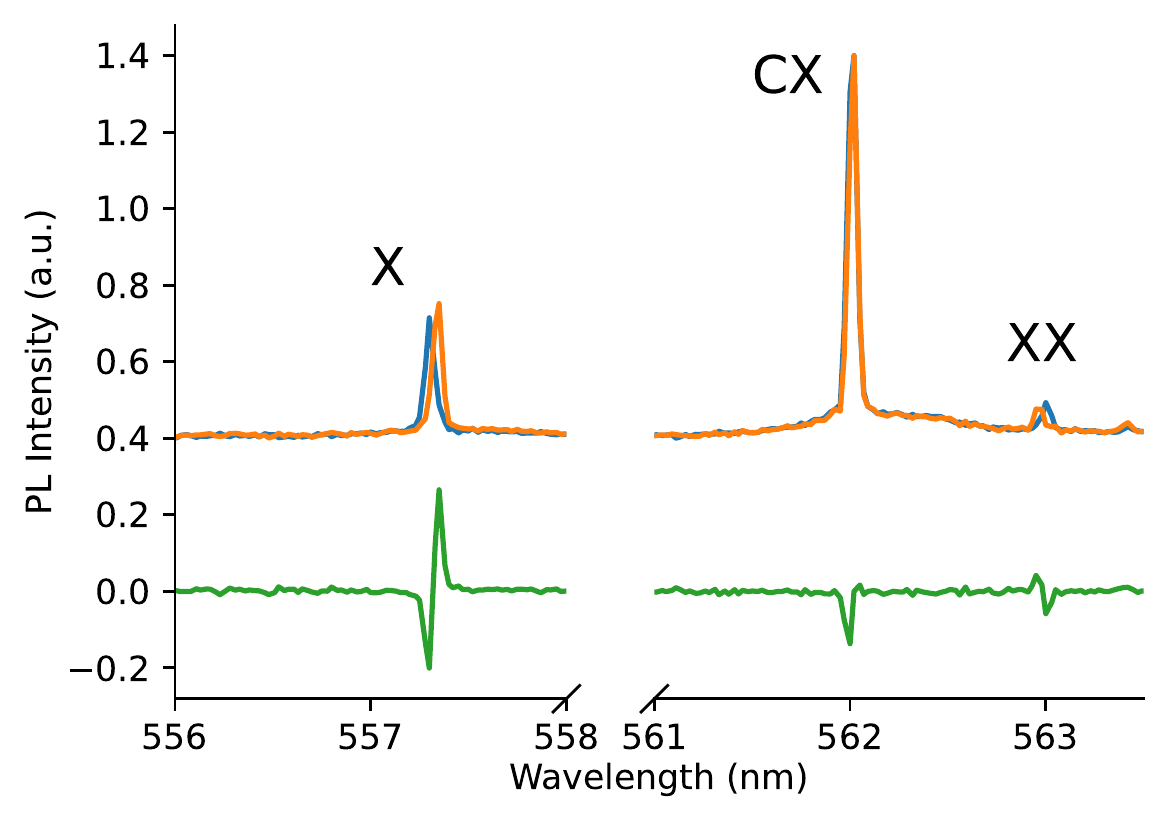}
\caption{Zoomed view of the three lines for two orthogonal linear polarizations. In that case, the excitation power is 10.5~\textmu W with a 405 nm CW laser. The green curve shows the subtraction between the two spectra.}
\label{fig:fine-structure}
\end{figure}

\vspace{0.5cm}
A fine-structure is observed with opposite signs for the lines assigned to X and XX as shown in Figure \ref{fig:fine-structure}. The central line does not show fine structure splitting, hence it is attributed to a charged exciton CX.

\subsection*{Low-temperature second order measurements}
\vspace{0.5cm}
Additionally, low-temperature autocorrelation measurements were performed on the studied nanowire and are presented in Figure \ref{fig:6K} below. 

\begin{figure}[htbp]
\centering
\includegraphics[width=1\linewidth]{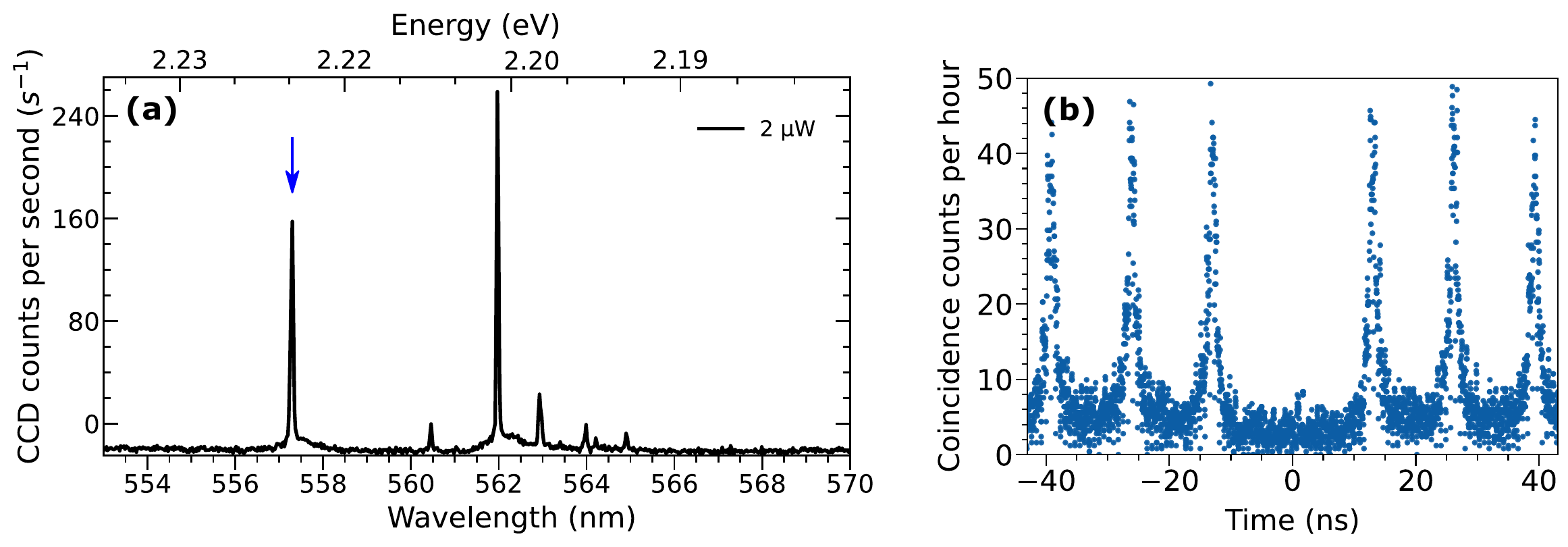}
\caption{\textbf{(a)} PL spectrum of the QD-NW taken at 6K under 2~\textmu W excitation power. \textbf{(b)} Second order correlation measurement performed on the line marked with a blue arrow in \textbf{(a)} (1-nm broad window).}
\label{fig:6K}
\end{figure}

\end{document}